\documentstyle[prb,aps,amsfonts,amssymb,epsfig]{revtex}
\begin{document}
\draft 
\twocolumn[\hsize\textwidth\columnwidth\hsize\csname @twocolumnfalse\endcsname
\title{Infrared signatures of the spin-Peierls transition in CuGeO$_3$}
\author{A. Damascelli and D. van der Marel}
\address{Solid State Physics Laboratory, University
 of Groningen, Nijenborgh 4, 9747 AG Groningen, The Netherlands}
\author{F. Parmigiani} 
\address{INFM and Dipartimento di Fisica, Politecnico 
di Milano, Piazza Leonardo da Vinci, 32-20133 Milano, Italy}
\author{G. Dhalenne and A. Revcolevschi} 
\address{Laboratoire de Chimie des Solides, Universit$\acute{e}$ de Paris-sud, 
B$\hat{a}$timent 414, F-91405 Orsay, France}
\date{\today}
\maketitle
\begin{abstract}
We investigated the infrared  reflectivity  of several
  Mg- and Si-substituted CuGeO$_3$ single crystals. The  temperature dependent
   {\em b}-axis and {\em c}-axis optical response is reported. For T$<$T$_{\rm{SP}}$ we detected 
   the activation of  zone-boundary phonons along the {\em b} axis of the crystal on the pure sample
    and for 1\% Mg and 0.7\% Si concentrations. From a detailed analysis of the phonon parameters the 
    redshift of the $B_{2u}$ mode at 48 cm$^{-1}$ is observed and discussed in relation to the soft 
    mode expected to drive the  spin-Peierls phase transition in CuGeO$_3$. Moreover, the polarization 
    dependence of a magnetic excitation  measured in  transmission at  44 cm$^{-1}$  has 
    been investigated.
\end{abstract}
%
%\pacs{PACS numbers: 63.20.Dj, 64.60.-i, 78.30.Hv}
\vskip2pc]
\narrowtext
In 1993 CuGeO$_3$ has been recognized, on the basis of magnetic susceptibility measurements,  \cite{hase} as the first inorganic compound showing a spin-Peierls (SP) transition , {\em i.e.}, a 
lattice distortion 
(due to the magneto-elastic coupling between the one-dimensional spin system and the three-dimensional 
phonon system) which occurs together with the formation of a spin-singlet ground state and the opening
 of a finite energy gap in the magnetic excitation spectrum. This discovery has renewed the interest in 
 the field of the SP phase transition, observed for the first time on an organic material in the 70's,  
 \cite{bray}  because the availability of large high quality single crystals of  CuGeO$_3$ makes it 
 possible to investigate this phenomenon by a very broad variety of experimental techniques. Furthermore, 
 CuGeO$_3$ seems to be a good candidate for the observation of a soft mode in the phonon spectrum, upon 
 passing through the SP transition. In fact, a well-defined soft mode  is expected in those theoretical 
 models describing an SP system in terms of a linear coupling between lattice and magnetic degrees of 
 freedom.  \cite{bulaevskii,cross} During the last few years both the structural deformation and the spin
  gap have been characterized in detail by x-ray and neutron scattering experiments.  \cite{hirota,lumsden,braden,fujita,ain} So far  no soft mode has yet been detected in CuGeO$_3$.

In this paper we present a detailed spectroscopic study of the vibrational and electronic signatures 
of the SP transition in pure and doped CuGeO$_3$.  We concentrate on the nature of the transition 
and on the dynamical interplay  between spins and phonons.  

We investigated the far- and mid-infrared  reflectivity (20$\,$-$\,$6000 cm$^{-1}$) of several 
 Cu$_{1-\delta}$Mg$_{\delta}$GeO$_3$  ($\delta$=0,$\,$0.01) and 
CuGe$_{1-x}$Si$_{x}$O$_3$ ({\em x}=0,$\,$0.007,$\,$0.05,$\,$0.1)
single crystals. These high-quality single 
crystals were grown 
from the melt by a floating zone technique. \cite{revcolevschi} Samples with dimensions of  approximately 1$\times$3$\times$6 mm$^3$ were aligned by conventional Laue diffraction 
and mounted in a liquid He flow 
cryostat to study the temperature dependence of the optical properties between 
4 K and 300 K. The reflectivity measurements were  performed with a Fourier transform 
spectrometer (Bruker IFS 113v), operating in near normal incidence configuration with polarized
 light in order 
to probe the optical response of the crystals along the {\em b} and the {\em c}  axes. The absolute 
reflectivities 
were obtained by calibrating the data acquired on the samples against a gold mirror.

The number and the symmetry of the infrared active phonons expected for the high temperature undistorted
 phase and the low temperature SP phase of  CuGeO$_3$ can be obtained from a group theoretical analysis 
 of the lattice vibrational modes. At room temperature CuGeO$_3$ has an orthorhombic crystal structure
  with lattice parameters
 {\em a}=4.81 \AA, {\em b}=8.47 \AA\    and {\em c}=2.941 \AA\      and space group
 {\em Pbmm} or,
 equivalently, {\em Pmma} in standard setting. \cite{vollenkle} The building blocks of the structure are
 edge-sharing CuO$_6$ octahedra and corner-sharing GeO$_4$ tetrahedra stacked along the 
{\em c} axis of the crystal and resulting in Cu$^{2+}$ and Ge$^{4+}$ chains parallel to the {\em c}  axis.
 These 
chains are linked together via the O atoms  [denoted as O(2)] and form layers parallel to the 
{\em b-c}  plane weakly coupled along the {\em a} axis. The irreducible representations of the optical
 vibrations of CuGeO$_3$,  in setting {\em Pbmm},  for T$>$T$_{\rm{SP}}$ is: \cite{ascona} 
\\
\\
$\Gamma=4A_{g}(aa,bb,cc)+4B_{1g}(ab)
+3B_{2g}(ac)+B_{3g}(bc)+3B_{1u}({\rm{E}}\|c)+
5B_{2u}({\rm{E}}\|b)+5B_{3u}({\rm{E}}\|a)~$,\\
\\
corresponding  to an expectation of 12 Raman active modes ($4A_{g}+4B_{1g}
+3B_{2g}+B_{3g}$) and 13 infrared active modes ($3B_{1u}
+5B_{2u}+5B_{3u}$). Below T$_{\rm{SP}}$ the crystal structure is still orthorhombic,
 but with lattice parameters a'=2$\times$a, {\em b'=b} and c'=2$\times$c and space group 
{\em Bbcm}  or , equivalently, {\em Cmca} in standard setting. \cite{hirota,braden} The distortion of
 the lattice in the phase transition can be characterized as the dimerization of the Cu-Cu pairs along 
 the {\em c} axis (dimerization out of phase in neighboring chains), together with the rotation of the
  GeO$_4$ tetrahedra around the axis defined by the O(1) sites (rotation opposite in sense for neighboring
   tetrahedra). Moreover, the O(2) sites of the undistorted structure split in an equal number of 
   O(2{\em a}) and O(2{\em b}) sites, distinguished by the distances O(2{\em a})-O(2{\em a}) and 
   O(2{\em b})-O(2{\em b}) shorter and larger than O(2)-O(2), \cite{braden} respectively. The irreducible
    representations of the optical vibrations,  in setting {\em Bbcm},  for T$<$T$_{\rm{SP}}$   
    is: \cite{ascona}
\\
\\
$\Gamma_{\rm{SP}}=8A_{g}(aa,bb,cc)+9B_{1g}(ab)+
7B_{2g}(ac)+6B_{3g}(bc)+5B_{1u}({\rm{E}}\|c)+
9B_{2u}({\rm{E}}\|b)+8B_{3u}({\rm{E}}\|a)~$.\\
\\
Therefore 30 Raman active modes ($8A_{g}+9B_{1g}+7B_{2g}+6B_{3g}$) and 22 infrared active modes 
($5B_{1u}+9B_{2u}+8B_{3u}$) are expected  for CuGeO$_3$ in the SP phase, all the additional vibrations
 being zone boundary modes activated by the folding of the Brillouin zone. In particular, the number 
 of infrared active phonons  is expected to increase from 5 to 8,  5 to 9 and 3 to 5 for  light 
 polarized along the {\em a}, {\em b} and {\em c} axis, respectively, upon passing through the phase 
 transition.

The {\em c}- and {\em b}-axis reflectivity spectra of pure CuGeO$_3$  are plotted in Fig.~1,  
for T=15 K (circles) and T=4 K (solid line). The data, characteristic of an ionic insulating material, are shown up to 1000 cm$^{-1}$ which covers  
the full phonon spectrum.  For 
T$>$T$_{\rm{SP}}$ three phonons are detected  along the {\em c} axis ($\omega_{\rm{TO}}\approx$\,167, 
528 and 715 cm$^{-1}$ for T=15 K), and five along the {\em b} axis ($\omega_{\rm{TO}}\approx$\,48, 210, 
286, 376 and 766  cm$^{-1}$ for T=15 K), in agreement with the theoretical expectation. The structure in 
Fig.~1(a) between 200 and 400 cm$^{-1}$ is due to a leakage of the modes polarized along the {\em c} axis. 
The feature at approximately 630 cm$^{-1}$  in Fig.~1(b) is a leakage of a mode polarized along the {\em a} 
axis. \cite{popovic} Whereas for E$\|c$ the spectra are exactly identical, a new feature is detected in the 
SP phase  at 800 cm$^{-1}$  for E$\|b$, as shown in the inset of  Fig.~1(b).  A careful 
investigation for temperatures ranging from 4 to 15 K (see Fig.~2) clearly shows that this feature, 
that falls in the frequency region of high reflectivity for the $B_{2u}$ phonon at 766 cm$^{-1}$ and
 therefore shows up in reflectivity mainly for its absorption, is activated by the SP transition. It
  corresponds to a new absorption peak in conductivity, superimposed on a background due to the 
  lorentzian tail of the close $B_{2u}$ mode (see inset of Fig.~2). We observed the same peak (at 
  the same resonant frequency) also on 1\% Mg and 0.7\% Si doped single crystals,  but not on 5\% 
  and 10\% Si doped samples, \cite{ascona} where we did not find any sign of the SP transition also 
  on the basis of  magnetic susceptibility measurements.

By fitting the reflectivity spectra with Lorentz oscillators for the optical phonons it is possible 
to obtain the temperature dependence  of the oscillator strength for the 800 cm$^{-1}$ feature. 
The results are plotted in 
Fig.~3, together with the peak intensity of the superlattice reflection measured by Harris {\em et al.}
 \cite{harris} in an x-ray scattering experiment on a pure CuGeO$_3$  single crystal  characterized, 
 as our sample, by T$_{\rm{SP}}$$\approx$13.2 K. From the perfect agreement of the infrared and x-ray
  scattering results on pure CuGeO$_3$ and from the observation that the resonant frequency is not 
  shifting at all with temperature,  we can conclude that the peak at 800 cm$^{-1}$ corresponds to a pure
   lattice excitation and the oscillator strength is proportional to the symmetry-breaking displacement of the atoms squared. The same conclusions can be obtained for the 1\% Mg and 0.7\% Si doped samples, 
   where T$_{\rm{SP}}$ of approximately 12.4 K and 9.3 K, respectively, are observed (in these two cases 
   the data can be compared, for 0.7\% Si doping, to those presented in Ref.~15 and , for 1\% Mg 
   doping, to those reported in Ref.~16 for a 0.9\% Zn-doped sample, because no x-ray nor neutron 
   scattering  data are available in the literature for Mg-doped CuGeO$_3$). Moreover, as this activated
    line does not show any frequency shift as a function of Si and Mg doping, we can conclude that it has to be a 
    folded zone boundary mode related to the $B_{2u}$  phonon observed at  766 cm$^{-1}$, which is mainly 
    an oxygen vibration. \cite{popovic} This gives for the $B_{2u}$ mode  an energy dispersion, over the 
    full Brillouin zone, of the order of 
24 cm$^{-1}$=2.98 meV at T=15 K. 

The reasons for not observing in our reflectivity spectra for T$<$T$_{\rm{SP}}$ all the phonons predicted
 from the group theoretical analysis, are probably the small values of the atomic displacements involved 
 in the SP transition ( with a correspondingly small oscillator strength of zone boundary modes), and/or 
 possibly the small dispersion of the optical branches of some of the lattice vibrations. However, it is
  not surprising that the only activated mode has been detected along the {\em b} axis of the crystal. In 
  fact, it is for this axis that for T$<$T$_{\rm{SP}}$  a spontaneous thermal contraction has been observed, 
  \cite{harris1} which can be responsible for a relative increase of the oscillator strength of the phonons
   polarized along the  {\em b} axis with respect to those polarized along the {\em c} axis.

In Fig.~3 the results of a fit of the experimental data by the equation 
 (1-T/T$_{\rm{SP}})^{2\beta}$ over a broad temperature range are also plotted,  for the pure sample.
 We obtained $\beta=0.26\pm0.02$  in agreement with Ref.~17. However, the best fit
  value of  $\beta$  is strongly dependent on the temperature range chosen to fit the data. If only 
points very close  (within one Kelvin)  to T$_{\rm{SP}}$  are considered,  the value
   $\beta=0.36\pm0.03$ is obtained, as reported in Ref.~6. It is not possible to perform 
   the same fit for the data acquired on doped samples because they are characterized by an upturned 
   curvature near T$_{\rm{SP}}$, which can be explained in terms of a distribution of transition 
   temperatures due to the disorder introduced upon doping the system. \cite{regnault}

As far as the dynamical interplay between spins and phonons in CuGeO$_3$  is concerned, it is clear
 from the reflectivity spectra plotted in Fig.~1 that a well-defined soft mode, driving the structural 
 deformation in CuGeO$_3$,  has not been detected in our measurements. However, interesting 
 information can be drawn from the temperature dependence of the phonon parameters obtained from the fit
  of the reflectivity data for the pure sample. In Fig.~4 the frequency shift (in per cent) for 
  the $B_{1u}$  modes (E$\|c$) and the $B_{2u}$ modes (E$\|b$) is plotted as a function of 
  temperature. We can clearly observe that the $B_{2u}$  mode at 48 cm$^{-1}$ 
  is the only one showing an evident monotonic redshift from 300 K to 15 K . This can be 
understood in terms of the normal-mode 
   displacements obtained by Popovi\'c {\em et al.} \cite{popovic} via a shell-model lattice dynamical
    calculation. In fact, this particular $B_{2u}$ mode, as well as the Raman active $B_{1g}$ at 
    116 cm$^{-1}$, consists mainly of the rotation (accompanied by a slight internal distortion) of
     the GeO$_4$ tetrahedra around the axis defined by the O(1) sites. It is precisely this 
``hinge motion''  \cite{khomskii} represented in terms of normal-mode displacements at {\em k}=($\pi/a$,0,$\pi/c$) which, together with the dimerization of the Cu-Cu pairs along the {\em c} axis, 
 corresponds to the structural deformation involved in the SP phase transition. The $B_{2u}$ mode 
at  48 cm$^{-1}$ has approximately the same character. Below T$_{\rm{SP}}$ this mode shows only a small blueshift. More noticeable there is a large ($\sim$15\%) reduction of oscillator strength
 {\em S} (see inset of  Fig.~4). One may speculate at this point, that a full softening is also absent 
for the phonons at {\em k}=($\pi/a$,0,$\pi/c$), implying that the phase transition is not driven 
by a softening of the phonon spectrum at 
{\em k}=($\pi/a$,0,$\pi/c$), but by a change in electronic structure which in turn determines the dynamical charge of the ions and the interatomic force constants. In this scenario the large change in 
oscillator strength of some of the vibrational modes results from a change in ionicity, or, in other words, 
a transfer of spectral weight from the elastic degrees of freedom to electronic excitations.

A last remark has to be made regarding the temperature dependence of the $B_{2u}$ mode ($B_{3u}$ 
in standard setting) observed at 286 cm$^{-1}$. In a recent paper \cite{li} a softening of this 
phonon, upon going through the phase transition, was suggested. This is not confirmed by our results
 which show no considerable frequency shift for this resonance upon reducing the temperature from 
 15 K to 4 K (see Fig.~4). On the other hand a reduction of both the scattering rate and the oscillator 
 strength is observed, which can explain the double-peak structure in the reflectance ratio R(20 K)/R(5 K)
  reported in Ref.~19.

Usually in optical spectroscopy direct singlet-triplet excitations are not detectable or very weak. 
However, such a transition has been observed 
 at 44.3 cm$^{-1}$ in an infrared transmission experiment where the singlet-triplet nature of the 
 transition was demonstrated by the Zeeman splitting observed in magnetic field. \cite{loosdrecht1} 
 In a very recent paper \cite{uhrig} this line was interpreted as a magnetic excitation across the 
 gap at the wave vector (0,$\pi/b$,0) in the Brillouin zone, activated by the existence of staggered
  magnetic fields along the {\em b} axis. In order to study this interpretation we measured the far infrared 
  transmission on the pure CuGeO$_3$ single crystal for E$\|b$ and E$\|c$. The absorbance difference spectra are reported in Fig.~5, where
    they have been shifted for clarity. For E$\|c$  no absorption is observed. However, for
     E$\|b$ an absorption peak, showing the appropriate temperature dependence, is present at 
     approximately 44 cm$^{-1}$. One has to note that the feature at 
     48 cm$^{-1}$ is produced by the low energy $B_{2u}$ phonon. In fact, due to the strong temperature
      dependence of its parameters, this line does not cancel out completely in the ratios of transmission
       spectra measured at different temperatures. The observed  polarization dependence 
 puts a strong experimental 
constraint on the possible microscopic mechanism giving rise to the singlet-triplet absorption peak.

In conclusion, we have investigated the temperature dependent phonon spectrum of pure and doped 
CuGeO$_3$, by means of  infrared reflectivity measurements. For T$<$T$_{\rm{SP}}$ we observed the
 activation of  zone-boundary phonons along the {\em b} axis of the crystals and the redshift of 
  the $B_{2u}$ mode at 48 cm$^{-1}$. The latter result has been  discussed in relation to the role
   played by this lattice vibration in driving the system into the dimerized phase. Moreover, a 
   magnetic excitation has been measured in transmission at  44 cm$^{-1}$ and its
    polarization dependence investigated.

We gratefully acknowledge M. Mostovoi and D.I. Khomskii 
for stimulating discussions and T.T.M. Palstra for the magnetic susceptibility measurements. We thank 
P.H.M. van Loosdrecht and  M. Gr\"uninger  for many useful comments. This investigation was supported by the Netherlands Foundation for
Fundamental Research on Matter (FOM) with financial aid from
the Nederlandse Organisatie voor Wetenschappelijk Onderzoek (NWO).

\begin{figure}[htb]
\centerline{\epsfig{figure=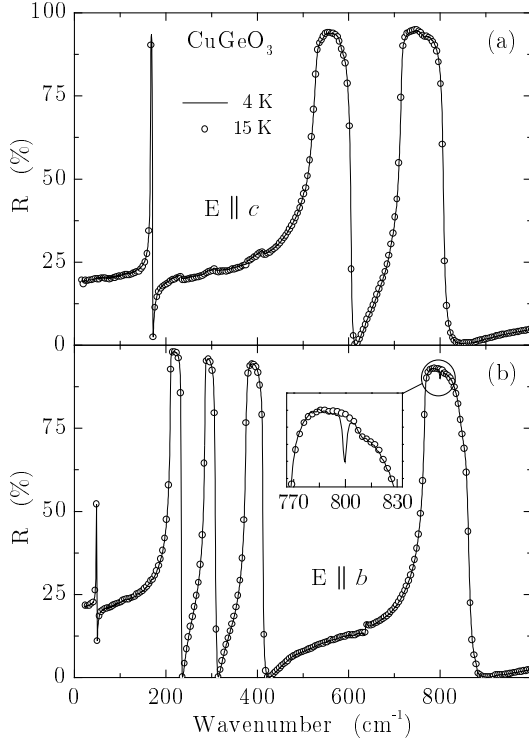,width=7cm,clip=}}
 \caption{Comparison between reflectivity spectra measured in the SP phase at 4 K 
(solid line) and just before the SP transition at 15 K (circles) on a pure single-crystal of
CuGeO$_3$. For light polarized along the {\em c} axis (a) no difference is found across the phase
 transition whereas for  light polarized along the {\em b} axis (b) a new feature appears 
at 800 cm$^{-1}$ (as clearly shown in the inset).
}
\label{fig1}
\end{figure}

\begin{figure}[htb]
\centerline{\epsfig{figure=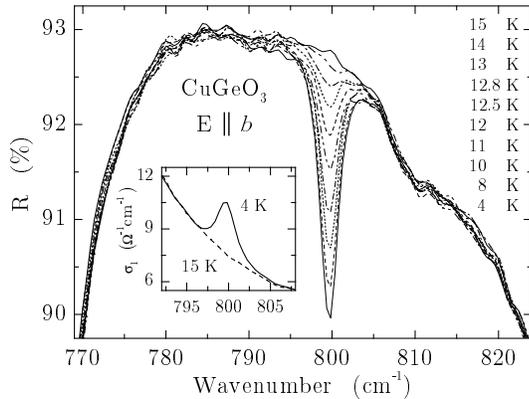,width=7cm,clip=}}
 \caption{Detailed temperature dependence of the feature observed with E$\|b$ in the reflectivity 
 spectra at 800 cm$^{-1}$  for T$<$T$_{\rm{SP}}$.  In the inset, where the dynamical conductivity 
 calculated via Kramers-Kronig analysis is plotted, a new peak is clearly visible for T=4 K.
}
\label{fig2}
\end{figure}

\begin{figure}[htb]
\centerline{\epsfig{figure=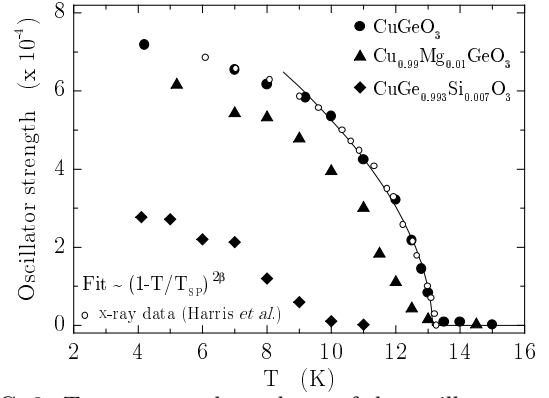,width=7cm,clip=}}
 \caption{Temperature dependence of the oscillator strength of  the zone boundary  mode observed
   for E$\|b$ at 800 cm$^{-1}$ on pure (filled circles), 1\% Mg-doped (triangles) and 0.7\% 
Si-doped (diamonds) CuGeO$_3$ single crystals. For comparison the x-ray scattering data of 
Harris {\em et al.} (Ref.~14) are also plotted (open circles).
}
\label{fig3}
\end{figure}

\begin{figure}[htb]
\centerline{\epsfig{figure=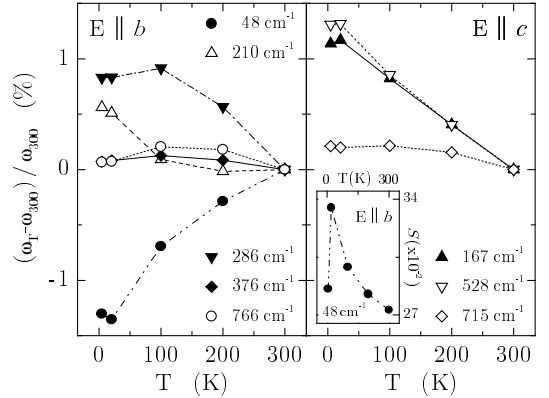,width=7cm,clip=}}
 \caption{Temperature dependence of the frequency shift (in per cent) of the transverse 
 optical phonons observed    for E$\|b$ and E$\|c$ on the pure single-crystal of  CuGeO$_3$. In 
the inset the oscillator strength {\em S} of the mode detected at 48 cm$^{-1}$,  for E$\|b$,  is plotted 
versus the temperature.
}
\label{fig4}
\end{figure}

\begin{figure}[htb]
\centerline{\epsfig{figure=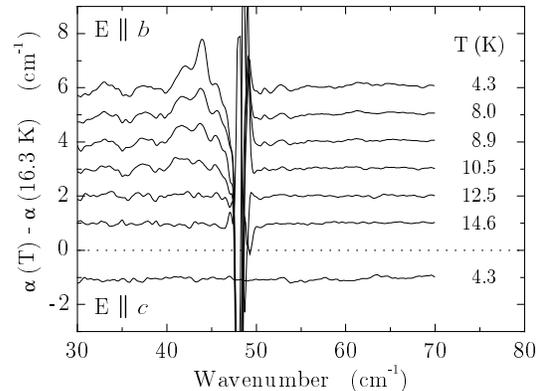,width=7cm,clip=}}
 \caption{Absorbance difference spectra  for the pure single-crystal of  CuGeO$_3$, with  
 E$\|b$ and E$\|c$. The spectra have been shifted for clarity.
}
\label{fig5}
\end{figure}

\end{document}